\documentclass[conference]{IEEEtran}
\IEEEoverridecommandlockouts
\usepackage{cite}
\usepackage{svg}
\usepackage{amsmath,amssymb,amsfonts, epsfig,latexsym}
\usepackage{algorithmic}
\usepackage{subfigure}
\usepackage{flushend}
\usepackage{hhline}
\usepackage{colortbl}
\usepackage{xcolor}
\usepackage{pdfpages}
\usepackage{graphicx}
\usepackage{textcomp}

\usepackage{multirow}
\usepackage{bigstrut}
\usepackage{array}
\usepackage{comment}
\usepackage{booktabs}
\usepackage{multicol}
\usepackage[utf8]{inputenc}
\renewcommand{\arraystretch}{1.5}

\def\BibTeX{{\rm B\kern-.05em{\sc i\kern-.025em b}\kern-.08em
    T\kern-.1667em\lower.7ex\hbox{E}\kern-.125emX}}

\usepackage[none]{hyphenat}
\usepackage[hyphens]{url}

\usepackage[textsize=tiny,textwidth=0.72in]{todonotes}
\usepackage{color} 
\usepackage{soul}

\begin{document}

\title{Accelerating LSTM-based High-Rate Dynamic System Models\\}

\author{\IEEEauthorblockN{Ehsan Kabir$^{\mathrm{*}}$, Daniel Coble$^{\mathrm{\$}}$, Joud N. Satme$^{\mathrm{\$}}$, Austin R.J. Downey$^{\mathrm{\$}}$, Jason D. Bakos$^{\mathrm{\dag}}$,\\ David Andrews$^{\mathrm{*}}$, Miaoqing Huang$^{\mathrm{*}}$}

\IEEEauthorblockA{\textit{$^{\mathrm{*}}$Department of Computer Science and Computer Engineering, University of Arkansas, USA} \\  
\textit{$^{\mathrm{\$}}$Department of Mechanical Engineering, University of South Carolina, USA}\\ 
\textit{$^{\mathrm{\dag}}$Department of Computer Science and Engineering, University of South Carolina, USA}\\ 
\{ekabir, dandrews, mqhuang\}@uark.edu, \{dncoble, jsatme\}@email.sc.edu, austindowney@sc.edu, jbakos@cse.sc.edu}}{\centering}





\onecolumn
{\large\vspace*{\fill}

© 2023 IEEE.  Personal use of this material is permitted.  
Permission from IEEE must be obtained for all other uses, 
in any current or future media, including reprinting/republishing this 
material for advertising or promotional purposes, 
creating new collective works, for resale or redistribution to servers or lists, 
or reuse of any copyrighted component of this work in other works. \\

This work has been accepted at the 
2023 33rd International Conference on Field-Programmable Logic and Applications (FPL)
and will appear in the proceedings and on the IEEE website soon.

\vspace*{\fill}
}
\twocolumn

\maketitle

\begin{abstract}
In this paper, we evaluate the use of a trained Long Short-Term Memory (LSTM) network as a surrogate for a Euler–Bernoulli beam model, and then we describe and characterize an FPGA-based deployment of the model for use in real-time structural health monitoring applications. The focus of our efforts is the DROPBEAR (Dynamic Reproduction of Projectiles in Ballistic Environments for Advanced Research) dataset, which was generated as a benchmark for the study of real-time structural modeling applications. The purpose of DROPBEAR is to evaluate models that take vibration data as input and give the initial conditions of the cantilever beam on which the measurements were taken as output.  DROPBEAR is meant to serve an exemplar for emerging high-rate ``active structures'' that can be actively controlled with feedback latencies of less than one microsecond.  Although the Euler–Bernoulli beam model is a well-known solution to this modeling problem, its computational cost is prohibitive for the time scales of interest. It has been previously shown that a properly structured LSTM network can achieve comparable accuracy with less workload, but achieving sub-microsecond model latency remains a challenge. Our approach is to deploy the LSTM optimized specifically for latency on FPGA. We designed the model using both high-level synthesis (HLS) and hardware description language (HDL). The lowest latency of 1.42 $\mu$S and the highest throughput of 7.87 Gops/s were achieved on Alveo U55C platform for HDL design.

\end{abstract}


\section{Introduction}
Long Short-Term Memory (LSTM) neural networks are capable of capturing dependencies for long sequential or temporal data in applications such as speech recognition, natural language processing, image captioning, scene analysis, etc. \cite{lstm1, lstm2}. In earlier works, such networks have been shown to be effective in utilizing time-series data to infer the state of a structure in a high-rate dynamic environment \cite{Nelson2022GeneratedDatasetsDynamic,high-rate4}.  The primary goal of this study is to develop a hardware-based LSTM model to enable ultra-low latency state estimation applications \cite{high-rate6}. 
High-rate dynamic systems refer to environments in which structures are subjected to impact loading that results in accelerations greater than 100 g for time periods of less than 100 ms \cite{Hong2018IntroductionStateEstimation}. Such systems are civil structures exposed to blasts, space infrastructures prone to debris strikes, and aerial vehicles capable of supersonic flight \cite{high-rate1,high-rate2}.  Systems exposed to such environments require rapid response, from event detection to decision-making, in the sub-millisecond or microsecond scale to ensure safe and reliable operations \cite{high-rate3,high-rate4,high-rate5}.

DROPBEAR data\cite{high-rate6, data} was used to come up with a suitable three-layer LSTM architecture that is trained off-line in software using logs recorded from the physical DROPBEAR apparatus. 
Our model showed an acceptable accuracy in terms of signal-to-noise ratio and desired latency on a real-time operating system (RTOS). However, the RTOS is unable to exploit the available parallelism available within LSTMs. On the other hand, FPGAs can accelerate inference with low power consumption using pipelined and parallel processing elements and are therefore suitable for LSTM implementation. Thus, a custom accelerator of the trained LSTM model was designed using both HLS and HDL in this paper.  Furthermore, an implementation on various FPGA platforms was carried out for performance enhancement and comparison. 
\hfill \\
\textit{The contributions of this paper are:}
\begin{itemize}
    \item [$\bullet$] Design of an LSTM accelerator framework using high-level synthesis (HLS) that meets the real-time requirements set by high-rate applications.  Results show that outermost loop pipelining generates a more efficient hardware design than outermost loop unrolling of the algorithm. 
    
    \item [$\bullet$] An alternative approach to the accelerator design using hardware description language (HDL) to improve performance. Results show that HDL provides the flexibility to choose the level of parallelism based on the available resources and timing requirements which not possible with the HLS-based approach. 
    
    \item [$\bullet$] An investigation into model deployment on several FPGA platforms from Xilinx to determine the best-performing configuration given the application. We targeted datacenter platforms such as Xilinx Alveo U55c and VC707, and an embedded platform, ZCU104. We found that the additional resources available in the U55C were unnecessary for the size of the deployed model. Nonetheless, the U55C's superior resources allowed maximum level of parallelism. Both ZCU104 and U55C boards achieve latency lower than VC707 because they achieve better frequency. U55C achieved the highest frequency of all, but its latency is lower than that of ZCU104 within the same level of parallelism.     
    
    
\end{itemize}


\section{Model Selection}\label{model}
\begin{figure}[h!]
\centering
\includegraphics[height=4.5cm, width=0.8\linewidth]{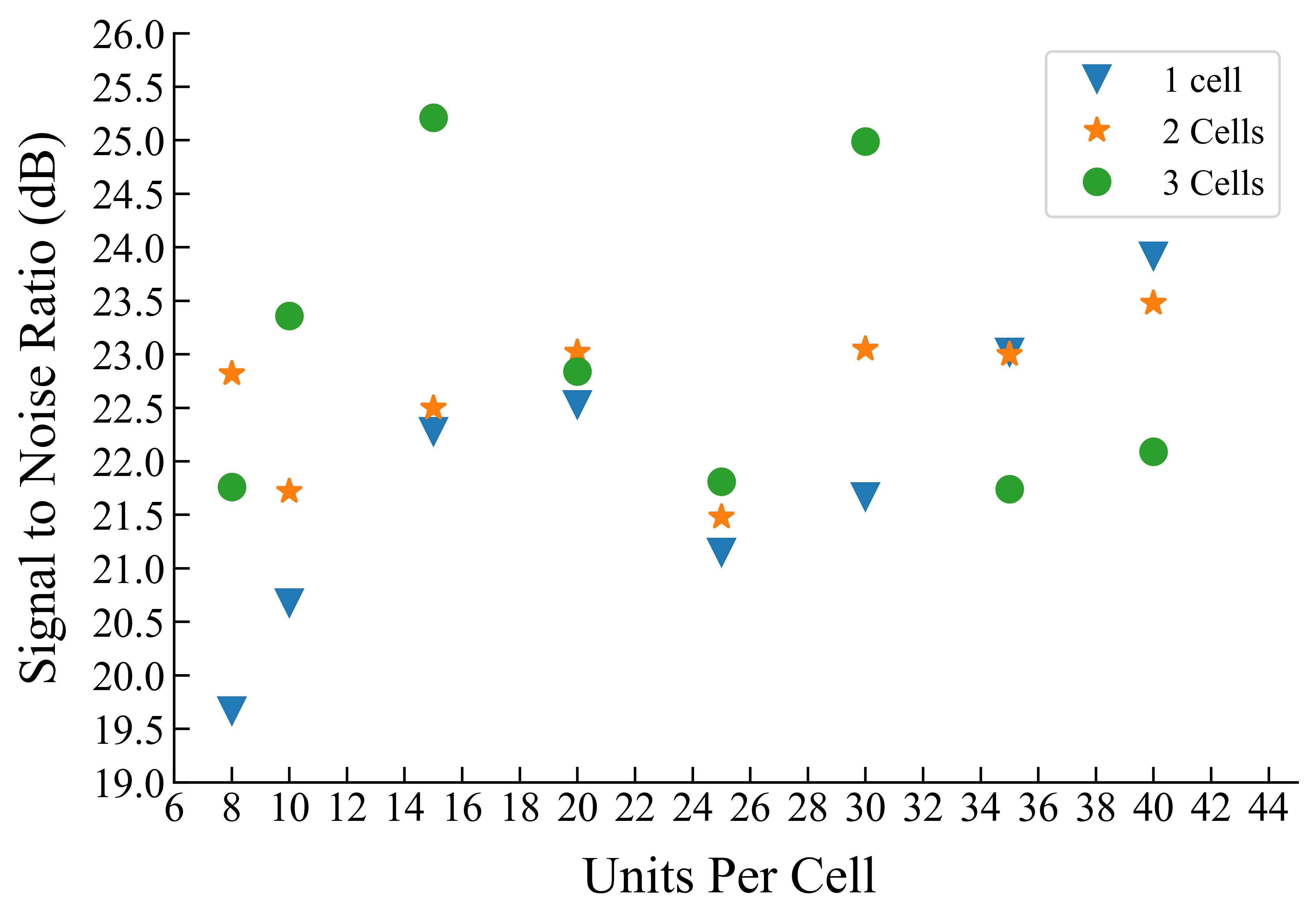}
\caption{\label{SNR}{SNR\textsubscript{dB} values of models with different numbers of cells and units}} 
\end{figure}

The LSTM model is chosen for a high-rate dynamic system to predict the real-time response within microsecond latency. The experimental data known as DROPBEAR was generated by Joyce et al.\cite{testbed}. The test setup and data are described in \cite{high-rate4, atiyeh}. 
Various LSTM models were trained on Python with Tensorflow and Keras to find a suitable model that meets the RTOS requirement of 500~$\mu$s in a real-time device consisting of a cRIO-9035 with a 1.33 GHz dual-core Intel Atom (E3825) manufactured by NI. 
The number of units per LSTM layer was varied from 8 to 40 units, each layer having the same number of units for simplicity. The number of layers was swept from 1 to 3. Fig.~\ref{SNR} shows a large variance in the measurement of signal-to-noise ratio (SNR) as we vary units per layer, though the SNR improves with increased number of layers. The 3-layer configuration with 15 units/layer is chosen in this paper for FPGA implementation as it has the highest SNR.
The model has 16 input features sourced from the input signal uniformly sampled across the previous timestep and produces an output state prediction every 500~$\mu$s on RTOS.
\section{Related work}\label{work}

LSTMs have been deployed in previous works using HLS \mbox{design}.
For example, work in \cite{relatedWork1} used HLS pragmas such as loop unrolling and pipelining to build a real-life speech recognition system using an LSTM model, whereas \cite{relatedWork2} implemented a real-time aircraft anomaly \mbox{detection system} using small scale LSTM on FPGA. A \mbox{multilayer} LSTM accelerator template was developed using the HLS tool for detecting gravitational waves which is a time-series data produced from LIGO detectors \cite{relatedWork3}. Low-power LSTM accelerators are built using pipeline and parallel algorithms in HLS \cite{relatedWork8}. Some works dealt with real-time response applications like electrical fault detection or heart rate monitoring with LSTM on FPGA after offline construction of a suitable LSTM architecture on python \cite{relatedWork4, relatedWork5}. 
Some HDL-based LSTM accelerators such as a human activity monitoring system described in \cite{relatedWork9}, provide flexibility of reconfiguration by adding parameterized features. The same group also detected artifacts from EEG signals by an LSTM model on FPGA at a low power and low frequency in \cite{relatedWork11}. Another research for healthcare applications reports an energy-efficient and high throughput real-time human action detection system \cite{relatedWork14}, where both HLS and HDL-based RTL modules are combined for the whole system design. 
Comparative studies between HLS and HDL-based designs have been done in the past for applications other than neural networks \cite{comparative1, comparative2}. Here, we compared the custom LSTM designs in both HDL and HLS formats.


\section{High Level Synthesis Implementation}\label{hls}

This section describes the high-level synthesis design technique. The core of the accelerator is designed in C++ language on Vitis HLS 2022.2.1 tool.

\begin{figure}[h]
\centering
\includegraphics[height=6cm, width=0.8\linewidth]{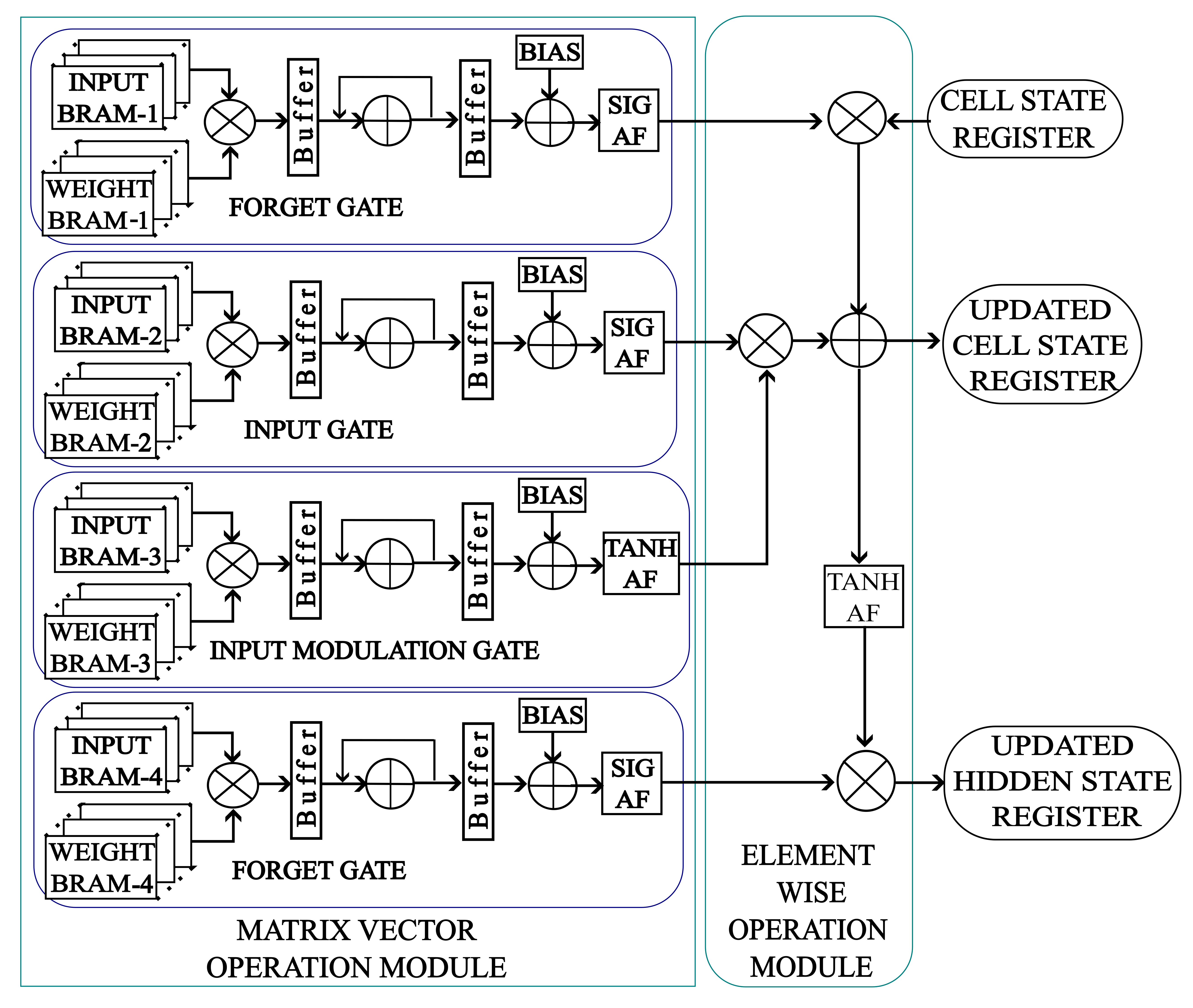}
\caption{\label{LSTMarch}LSTM Operations for High Level Synthesis Design.}
\end{figure}

There are two main units in the accelerator architecture - the matrix-vector operations (MVO) unit and the element-wise operations (EVO) unit. HLS design did not have any function defined for them. As a result, the exported RTL did not have separate RTL instances for them.
The LSTM gates within the MVO unit are separately defined as functions with unique arguments, allowing for the generation of independent parallel RTL modules. Instead of any distinct functions, the EVO unit's operations are expressed as distinct for-loops. Fig. \ref{LSTMarch} gives a high-level overview of the HLS implementation of the LSTM accelerator. Each gate can perform multiplication and accumulation (MAC) operations either in parallel or sequentially depending on the number of BRAMs.
Since each of the LSTM network layers function in succession, the gate modules are reused for operations in various layers.
Each gate module contains two for-loops.
One loop iterates over the hidden units. This loop contains two other distinct for-loops, one for multiplication and the other for accumulation operations. The other loop inside the function executes the summing operation with the bias over each hidden unit before executing the activation function. 
The main arguments of the gate functions are the inputs, hidden states, input weights, recurrent weights, and bias vectors. The array partition pragma entirely partitions the vectors of the inputs, hidden states, and biases to create registers and permit parallel access because the size of these vectors is small.
However, the vectors for input and recurrent weights are represented as BRAMs to store a large number of elements. Depending on the size of the LSTM network and the compiling capacity of the synthesis tool, they can be partially or entirely partitioned to generate multiple BRAMs. Large array partitioning slows down the compilation flow and synthesis process may even stop.
For an easier representation of the multiplication and accumulation operations of LSTM on HLS, the inputs and hidden state vectors, as well as the input and recurrent weight vectors, are concatenated. 
For fully pipelined operations, pipeline pragmas were applied on the outer loops of the functions, 
which fully unrolled the internal loops, but the operations were not fully parallel because of the BRAMs. The number of DSPs used for multiplication appears to depend on the size of the concatenated vector of inputs and hidden states. However, they do not start computation at the same clock cycle even being allocated simultaneously which is a limitation of HLS. 
To increase the utilization of DSPs for parallel multiplications, the outermost loop can be unrolled by some factors depending on the available resources. 
The EVO unit contains several for loops, but all loop operations are pipelined, and no loop was unrolled.
\section{Register Transfer Level Implementation}\label{rtl}
\begin{figure}[h]
\centering
\includegraphics[height=4cm, width=0.7\linewidth]{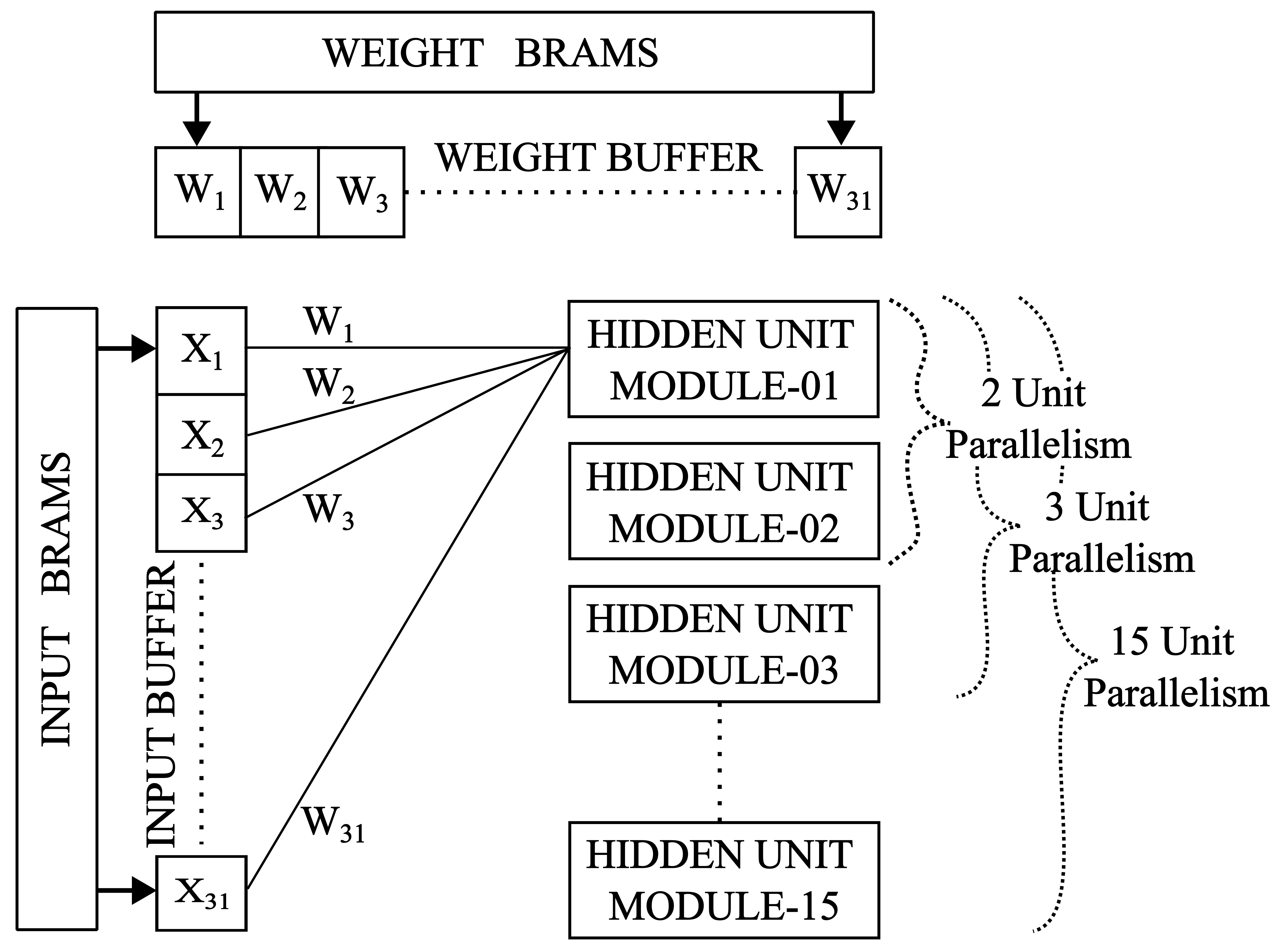}
\caption{\label{RTLarch}LSTM Operations for HDL Design}
\end{figure}

Since RTL provides added flexibility as compared to HLS, we developed a Verilog implementation of the LSTM and synthesized it with Vivado 2022.2.1 tool.  Fig.~\ref{RTLarch} shows the connections of modules in the RTL implementation.
 

The gate modules and the MVO module are not defined separately in the HDL design. It aided in reducing connections between modules that consume extra LUTs. For parallel execution, the hidden units within each gate are defined as modules and instantiated multiple times. This module contains some descriptions of logic operations as well as the instances of other modules such as a multiplier, adder, and activation functions. The number of this module to instantiate for each LSTM gate at the top module is configurable. It indicates a total number of parallel operations which is shown by unit parallelism in Fig.~\ref{RTLarch}. We were able to increase the parallel operations with parallel DSPs in this manner, which was not possible in HLS. HDL design required parallel DSPs for the EVO unit. 
The weights are stored in separate BRAMs in each hidden unit. As parallel DSPs require input data simultaneously, the number of BRAM instances grows in proportion to the number of hidden unit instances. The weights stored in the BRAMs are first transferred to the registers (w1, w2,...,w31 in Fig.~\ref{RTLarch}) to facilitate parallel data access. As the number of DSPs was increased, performance improved dramatically over the HLS design. Because of the heavy usage of DSPs, the design becomes crowded, preventing high-frequency operation. LUT usage rises so that correct data gets multiplexed to the DSPs. As the size of this LSTM is tiny, the number of concatenated inputs and hidden states were kept constant. Only flexibility over hidden units was demonstrated, but the same flexibility may be extended to inputs as well. 
\section{Overall System}\label{system}
\begin{figure}[h!]
\centering
\includegraphics[height=4cm, width=0.6\linewidth]{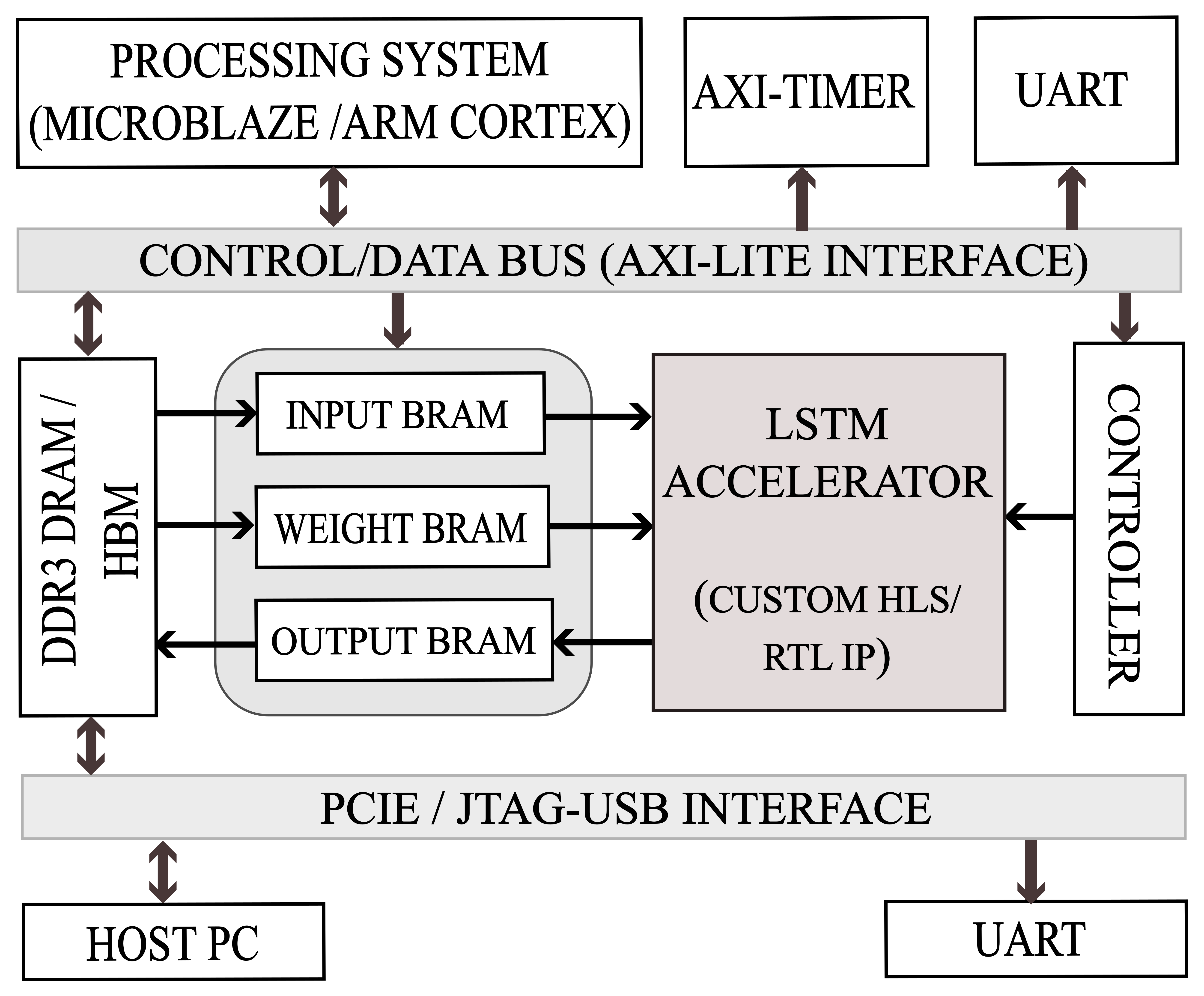}
\caption{\label{System}Complete System Design}
\end{figure}

Fig.~\ref{System} shows the complete system design for running the LSTM model on different FPGA platforms such as VC707 (Virtex-7 XC7VX485TFFG1761-2), ZCU104 (Zynq UltraScale+XCZU7EV-2FFVC1156 MPSoC) and U55C (UltraScale+XCU55C-FSVH2892-2L-E) for our experiments. The overall system was designed on Vivado 2022.1.2 design suite. It contains a custom IP block for the LSTM accelerator (LA), which can either be exported from HLS or be built directly with HDL. The LA has internal BRAMs as shown in Fig.~\ref{LSTMarch}. However, the system design has some external block ram generator modules for storing inputs, weights, and outputs. These inputs and weights are fetched from external DDR3 DRAM or High Bandwidth Memory (HBM). The outputs are returned to DRAM or HBM. Because of the limited local memory, the software executing on the CPUs is also saved on DRAM or HBM.


Both VC707 and ZCU104 boards have onboard DRAM memory, while the Alveo U55C contains HBMs.
The DRAM of VC707 is connected to the programmable logic (PL) side. It can communicate with the MicroBlaze ($\mu$B) softcore processing system (PS) using a memory interface generator (MIG) connected by AXI-lite interface. The $\mu$B is configured for the maximum frequency operation \cite{microblaze}.
On ZCU104, we access the onboard DRAM through the ARM-based Multiprocessor System-on-Chip (MPSoC) subsystem.
We used the same ARM processor to run our LSTM model and check the performance only in PS.
On U55c, the $\mu$B can access the HBMs. All the boards are connected to the same HOST PC with USB-JTAG interface. U55C was connected with a PCIe 3.0$\times$4 interface of another host which is a server with Intel Xeon CPU E5-2603 v4 @1.70GHz, so that it receives enough power to run and gets reset signal after power on. This host can communicate with other IPs except the PS using the DMA/Bridge Subsystem for PCI Express IP \cite{DMA} in the system. PS uses AXI-TIMER to measure the latency which includes the time between the start and the stop signal from the custom IP module. The host connected with JTAG cable displays the results on the terminal using the UARTlite interface. 
\section{Results and Evaluation}\label{results}


On three separate platforms with varying levels of parallelism, results for 32-bit (FP-32), 16-bit (FP-16), and 8-bit (FP-8) fixed point precision were obtained. Then they were compared in terms of maximum frequency (Fmax), resource utilization, latency, throughput, and normalized throughput \cite{CNN-MLPA}.

Fmax is reported for the system design in Fig.~\ref{System}, which includes all IPs, while resource usage is only reported for the accelerator. This is because the accelerator is identical across all platforms, but the overall system design differs among them. For a fair comparison, the bit lengths of the inputs, outputs, weights, and intermediate values are maintained the same in both HLS and HDL designs.
The findings for HLS design reveal that for the same pragmas such as loop unroll, loop pipeline, and array partition, the number of BRAM and DSP is different in different platforms and bit precision.
HLS tends to optimize itself regardless of the pragmas in different platforms. Hence, array partition was done with different factors on different platforms so that the number of DSPs remained the same. Despite the reduction in resource usage and increase in frequency for FP-8, the model did not automatically utilize the freed-up resources to decrease the delay. The improvement in frequency resulted in a minor reduction in latency. To further reduce latency for FP-8, the design must be updated again. The results in Table-\ref{loop} compares the effect of pipelining and unrolling the outermost loop inside each gate. Unrolling the outermost loop entirely or partially did not enhance performance significantly, even though resource use, such as DSP, increased by 8$\times$.
\begin{table}[htbp]
\setlength{\arrayrulewidth}{0.2mm}
\renewcommand{\arraystretch}{0.8}
  \centering
  \caption{HLS Loop Optimization}
  \resizebox{1.0\columnwidth}{0.8cm}{%
    \begin{tabular}{|c|c|c|c|c|}
    \toprule
    \multirow{2}[2]{*}{HLS Designs} & \multicolumn{1}{c|}{\multirow{2}[2]{*}{Platform \& Precision}} & \multicolumn{1}{c|}{\multirow{2}[2]{*}{\ \ DSP\ \ }} & \multicolumn{1}{c|}{Maximum} & \multicolumn{1}{c|}{Latency} \\
    \multicolumn{1}{|c|}{} &   \multicolumn{1}{c|}{}     &       &  \multicolumn{1}{c|}{Frequency (MHz)}     &    \multicolumn{1}{c|}{($\mu$S)} \\
    \midrule
    Loop Unroll & \multicolumn{1}{c|}{Virtex 7} & 1852  & 166   & 6.12 \\
\cmidrule{1-1}\cmidrule{3-5}    Loop Pipeline &   Fixed-16     & 224   & 250   & 6.54 \\
    \bottomrule
    \end{tabular}%
    }
  \label{loop}%
\end{table}%
\raggedbottom

On the other hand, the number of DSP can be controlled in HDL design. Parallelism is increased by increasing DSP blocks, however doing so leads to congestion in the routing system, which reduces overall frequency and occasionally results in no routing at all. As a result, it is important to carefully manage the amount of parallelism to avoid drastically decreasing frequency or going over the resource limit. 
Although the increment of DSP causes a reduction of frequency, the performance gets better than that of HLS designs. The number of DSP depends on the number of hidden units in HDL. In order to prevent the frequency from declining and resources from being overused, we reduced the number of parallelism as the bit width increased up to 32. The results in Table-\ref{HLS} and Table-\ref{HDL} also indicate that after 32-bit precision, HLS design starts performing better than the HDL design. It is because DSPs are heavily utilized in HDL design resulting in frequency decay. Furthermore, as precision rises, more DSPs for MAC are utilized, causing a resource overflow. As a result, HDL is unable to maximize parallelism for FP-32 without sacrificing frequency. Full parallelism can be achieved for our LSTM model up to 16-bit precision in all the FPGA platforms except ZCU104 which exceeds available DSPs if more than 2 unit parallelism is applied. With full parallelism (15 units for our model), U55C achieves the lowest latency of all as shown in Table-\ref{unit}. Thus, increasing the parallelism improved performance more than the HLS design in spite of the frequency drop.
\begin{table}[h]
\setlength{\arrayrulewidth}{0.05mm}
\renewcommand{\arraystretch}{0.8}
  \centering
  \caption{Effects of Parallelism on HDL Design}
  \resizebox{0.7\columnwidth}{1.2cm}{%
    \begin{tabular}{|c|c|c|c|c|c|c|}
    \toprule
    \multicolumn{1}{|c|}{\multirow{3}[2]{*}{Platform}} & \multicolumn{1}{c|}{\multirow{3}[-4]{*}{Bit}} & \multicolumn{1}{c|}{\multirow{3}[-4]{*}{LUT}} & \multicolumn{1}{c|}{\multirow{3}[-4]{*}{DSP}} & \multicolumn{1}{c|}{\multirow{3}[-12]{*}{Highest}} & \multicolumn{1}{c|}{\multirow{2}[-4]{*}{\cellcolor[rgb]{ 1,  .949,  .8}{Fmax}}} & \multicolumn{1}{c|}{\multirow{3}[-12]{*}{\cellcolor[rgb]{ 1,  .949,  .8}Latency}} \\
          & \multicolumn{1}{c|}{\multirow{3}[-4]{*}{Precision}}      &   \multicolumn{1}{c|}{\multirow{3}[-4]{*}{(\%)}}    &  \multicolumn{1}{c|}{\multirow{3}[-4]{*}{(\%)}}      &  \multicolumn{1}{c|}{\multirow{3}[-12]{*}{Level of}}     &   \multicolumn{1}{c|}{\multirow{3}[-6]{*}{\cellcolor[rgb]{ 1,  .949,  .8}(MHz)}}    &  \multicolumn{1}{c|}{\multirow{3}[-6]{*}{\cellcolor[rgb]{ 1,  .949,  .8}($\mu$S)}}\\
          &       &       &       &  \multicolumn{1}{c|}{\multirow{3}[-12]{*}{Parallelism}}     &  \multicolumn{1}{c|}{{\cellcolor[rgb]{ 1,  .949,  .8}{} }}    & \multicolumn{1}{c|}{{\cellcolor[rgb]{ 1,  .949,  .8}{} }} \\
    \midrule
    \multicolumn{1}{|c|}{\multirow{2}[4]{*}{Virtex 7}} & \multicolumn{1}{c|}{FP-32} & 28    & 69    & \multicolumn{1}{c|}{4 Units} & \cellcolor[rgb]{ 1,  .949,  .8}{142} & \cellcolor[rgb]{ 1,  .949,  .8}{5.78} \\
\cmidrule{3-7} \multicolumn{1}{|c|}{}         & \multicolumn{1}{c|}{FP-16} & 39    & 72    & \multicolumn{1}{c|}{15 Units} & \cellcolor[rgb]{ 1,  .949,  .8}166 & \cellcolor[rgb]{ 1,  .949,  .8}{2.06} \\
    \midrule
    \multirow{2}[4]{*}{U55C} & FP-32 & 11    & 38    & 8 Units & \cellcolor[rgb]{ 1,  .949,  .8}{150} & \cellcolor[rgb]{ 1,  .949,  .8}{2.38} \\
\cmidrule{3-7}          & FP-16 & 9     & 22    & 15 Units & \cellcolor[rgb]{ 1,  .949,  .8}{250} & \cellcolor[rgb]{ 1,  .949,  .8}{1.42} \\
    \bottomrule
    \end{tabular}%
    }
  \label{unit}%

\end{table}%
\raggedbottom

For HLS design of FP-8, DSPs were only employed for the activation functions because DSPs is not used below 10-bit precision. Although we compelled the use of DSPs for our multipliers in HDL design by employing Verilog attributes, their proper sharing could not be obtained which would have reduced consumption. Only the LUT and FF consumption was decreased by low bit precision. Thus, FP-8 will be useful for bigger models. One important improvement with FP-8 is achieving high frequency, and it helped reduce the latency to some extent. The total number of operations was determined for our LSTM model from which the throughput (Giga operations/second [GOPS]) was computed \cite{CNN-MLPA}. For a fair comparison between HLS and HDL-based design, normalized throughput both with respect to the LUTs (GOPS/LUT) and DSPs (GOPS/DSP) was calculated. Same parameters are also used for a fair comparison with different LSTM models in other works. 
HDL design has the flexibility to increase the resources to maximize throughput. Thus, throughput is higher than the HLS design in all the platforms. However, the (GOPS/LUT) and (GOPS/DSP) are higher in HLS design because it consumes fewer resources. As HLS automates the majority of optimization procedures, there is little scope to increase the resources to decrease latency. Table-\ref{HLS} reports all data related to the HLS design on different platforms. ZCU104 achieves the lowest latency, the highest GOPS, and the highest GOPS/LUT and GOPS/DSP for all precision. 
\begin{table}[h]
\setlength{\arrayrulewidth}{0.2mm}
\renewcommand{\arraystretch}{0.8}
  \centering
  \caption{Results for High-Level Synthesis Design}
    \resizebox{1.0\columnwidth}{1.6cm}{%

    \begin{tabular}{|c|c|c|c|c|c|c|c|c|c|c|}
    \toprule
    \multicolumn{1}{|c|}{\multirow{3}[2]{*}{Platform}} & \multirow{2}[1]{*}{Bit} & \multirow{3}[2]{*}{LUT} & \multirow{3}[2]{*}{FF} & \multirow{3}[2]{*}{BRAM 36k} & \multirow{3}[2]{*}{DSP} & \multirow{2}[1]{*}{{Fmax}} & {\multirow{2}[1]{*}{Latency}} & \multicolumn{1}{c|}{\multirow{2}[1]{*}{Throughput}} & \multicolumn{1}{c|}{\multirow{3}[-2]{*}{{GOPS/}}} & \multicolumn{1}{c|}{\multirow{3}[-2]{*}{{GOPS/}}} \\
          & \multicolumn{1}{c|}{\multirow{2}[2]{*}{Precision}} & \multicolumn{1}{c|}{} & \multicolumn{1}{c|}{} & \multicolumn{1}{c|}{} & \multicolumn{1}{c|}{} & \multirow{2}[1]{*}{(MHz)} &  \multicolumn{1}{c|}{\multirow{2}[2]{*}{($\mu$S)}}  &  \multicolumn{1}{c|}{\multirow{2}[2]{*}{(GOPS)}}     &  \multicolumn{1}{c|}{\multirow{3}[-2]{*}{{LUT}}}     &  \multicolumn{1}{c|}{\multirow{3}[-2]{*}{{DSP}}} \\
          & \multicolumn{1}{c|}{} & \multicolumn{1}{c|}{} & \multicolumn{1}{c|}{} & \multicolumn{1}{c|}{} & \multicolumn{1}{c|}{} &   &       &       &       &  \bigstrut[b]\\
    \midrule
    \multicolumn{1}{|c|}{\multirow{3}[2]{*}{Virtex 7}} & FP-32 & 70380 (23\%) & \multicolumn{1}{c|}{86579 (14\%)} & 41.5 (4\%) & 712 (25\%) & 210   & 8.75  & 1.28  & 18.19 & 1.80 \bigstrut[t]\\
          & FP-16 & 30532 (10\%) & 36186 (6\%) & 22 (2\%) & 224 (8\%) & 213   & 7.4   & 1.51  & 49.46 & 6.74 \\
          & FP-8 & 26889 (9\%) & 20683 (3\%) & 0 (0\%) & 30 (1\%) & 235   & 6.36  & 1.76  & 65.45 & 58.67 \bigstrut[b]\\
    \midrule
    \hline
    \multicolumn{11}{|c|}{}    \bigstrut[t]\\[-1.0em]
    \multicolumn{1}{|c|}{\multirow{3}[2]{*}{ZCU104}} & \multirow{3}[-6]{*}{FP-32} & \multirow{3}[-6]{*}{78850 (34\%)} & \multirow{3}[-6]{*}{94936 (21\%)} & \multirow{3}[-6]{*}{17.5 (16\%)} & \multirow{3}[-6]{*}{712 (41\%)} & \multirow{3}[-6]{*}{305}   & \multirow{3}[-6]{*}{3.74}  & \multirow{3}[-6]{*}{2.99}  & \multirow{3}[-6]{*}{37.92} & \multirow{3}[-6]{*}{4.20} \\
          & \multirow{3}[-6]{*}{FP-16} & \multirow{3}[-6]{*}{36458 (16\%)} & \multirow{3}[-6]{*}{39326 (9\%)} & \multirow{3}[-6]{*}{10 (3\%)} & \multirow{3}[-6]{*}{224 (13\%)} & \multirow{3}[-6]{*}{350}   & \multirow{3}[-6]{*}{2.92}  & \multirow{3}[-6]{*}{3.83}  & \multirow{3}[-6]{*}{105.05} & \multirow{3}[-6]{*}{17.10} \\
          & \multirow{3}[-6]{*}{FP-8} & \multirow{3}[-6]{*}{23575 (10\%)} & \multirow{3}[-6]{*}{21590 (5\%)} & \multirow{3}[-6]{*}{0 (0\%)} & \multirow{3}[-6]{*}{15 (1\%)} & \multirow{3}[-6]{*}{400}   & \multirow{3}[-6]{*}{2.83}  & \multirow{3}[-6]{*}{3.95}  & \multirow{3}[-6]{*}{167.55} & \multirow{3}[-6]{*}{263.33} \bigstrut[b]\\
          \midrule
          \hline
    \multicolumn{9}{|c|}{}                                                & \multicolumn{1}{c}{} &  \bigstrut[b]\\[-1.0em]
    \multicolumn{1}{|c|}{\multirow{3}[2]{*}{U55C}} & FP-32 & 64930 (5\%) & 80191 (3\%) & 29.5 (1\%) & 711 (8\%) & 362   & 6.86  & 1.63  & 25.10 & 2.29 \bigstrut[t]\\
          & FP-16 & 25346 (2\%) & 31136 (1\%) & 16 (1\%) & 224 (2\%) & 375   & 4.72  & 2.36  & 93.42 & 10.57 \\ 
          & FP-8 & 23899 (2\%) & 17422 (1\%) & 0 (0\%) & 15 (0.2\%) & 380   & 4.65  & 2.4  & 100 & 160.00 \bigstrut[b]\\ 
    \bottomrule
    \end{tabular}%
    }
  \label{HLS}%
\end{table}%
\raggedbottom

Table-\ref{HDL} reports all data related to the HDL design on different platforms for 2 unit parallelism. The utilization of LA is the same regardless of the platforms. ZCU104 shows the best performance among other platforms for HDL design also. The utilization is higher than HLS design, so latency was reduced by $1.34\times$. GOPS/LUT is close to HLS design, but GOPS/DSP is much lower because HDL design mainly reduces latency by parallel operations of DSPs.
\begin{table}[h]
\setlength{\arrayrulewidth}{0.2mm}
\renewcommand{\arraystretch}{0.8}
  \centering
  \caption{Results for Hardware Description Language Synthesis Design}
    \resizebox{1.0\columnwidth}{1.5cm}{%
    \begin{tabular}{|c|p{4.045em}|c|c|c|c|c|c|c|c|c|}
    \toprule
     \multicolumn{1}{|c|}{\multirow{2}[2]{*}{Platform}} & \multicolumn{1}{c|}{Bit}  & \multicolumn{1}{c|}{LUT} & \multicolumn{1}{c|}{FF} & \multicolumn{1}{c|}{BRAM} & \multicolumn{1}{c|}{DSP} & \multicolumn{1}{c|}{Fmax} & \multicolumn{1}{c|}{Latency} & \multicolumn{1}{c|}{Throughput} & \multicolumn{1}{c|}{GOPS/} & \multicolumn{1}{c|}{GOPS/} \\
        & \multicolumn{1}{c|}{Precision} &    \multicolumn{1}{c|}{(\%)}    &  \multicolumn{1}{c|}{(\%)}     &     \multicolumn{1}{c|}{36k (\%)}   &  \multicolumn{1}{c|}{(\%)} &   \multicolumn{1}{c|}{(MHz)}   &  \multicolumn{1}{c|}{($\mu$S)}     &    \multicolumn{1}{c|}{(GOPS)}    &  \multicolumn{1}{c|}{LUT}     &  \multicolumn{1}{c|}{DSP}\\
    \midrule

    \multicolumn{1}{|c|}{\multirow{3}[1]{*}{Virtex 7}} &  \multicolumn{1}{c|}{FP-32} & 17    & 16    & 1     & 43    & 150   & 11.48 & 0.97  & 19.34 & 0.81 \\
          &  \multicolumn{1}{c|}{FP-16} & 22    & 23    & 5     & 41    & 166   & 3.71  & 3.01  & 45.19 & 2.64 \\
          &  \multicolumn{1}{c|}{FP-8} & 13    & 12    & 5     & 35    & 200   & 3.10  & 3.61  & 95.06 & 3.64 \\
    \midrule
    \hline
    \multicolumn{1}{|c|}{\multirow{3}[2]{*}{ZCU104}} &  \multicolumn{1}{c|}{\multirow{2}[-2]{*}{FP-32}} & \multirow{2}[-2]{*}{22}    & \multirow{2}[-2]{*}{21}    & \multirow{2}[-2]{*}{4}     & \multirow{2}[-2]{*}{69}    & \multirow{2}[-2]{*}{230}   & \multirow{2}[-2]{*}{7.11}  & \multirow{2}[-2]{*}{1.57}  & \multirow{2}[-2]{*}{31.62} & \multirow{2}[-2]{*}{1.31} \\
          &  \multicolumn{1}{c|}{\multirow{2}[-2]{*}{FP-16}} & \multirow{2}[-2]{*}{30}    & \multirow{2}[-2]{*}{29}    & \multirow{2}[-2]{*}{15}    & \multirow{2}[-2]{*}{66}    & \multirow{2}[-2]{*}{250}   & \multirow{2}[-2]{*}{2.14}  & \multirow{2}[-2]{*}{5.21}  & \multirow{2}[-2]{*}{76.69} & \multirow{2}[-2]{*}{4.56} \\
          &  \multicolumn{1}{c|}{\multirow{2}[-2]{*}{FP-8}} & \multirow{2}[-2]{*}{16}    & \multirow{2}[-2]{*}{16}    & \multirow{2}[-2]{*}{15}    & \multirow{2}[-2]{*}{57}    & \multirow{2}[-2]{*}{300}   & \multirow{2}[-2]{*}{1.72}  & \multirow{2}[-2]{*}{6.50}  & \multirow{2}[-2]{*}{171.61} & \multirow{2}[-2]{*}{6.55} \\
    \midrule
    \hline
    \multicolumn{1}{|c|}{\multirow{3}[2]{*}{U55C}} &  \multicolumn{1}{c|}{\multirow{2}[-2]{*}{FP-32}} & \multirow{2}[-2]{*}{4}    & \multirow{2}[-2]{*}{4}     & \multirow{2}[-2]{*}{1}     & \multirow{2}[-2]{*}{13}    & \multirow{2}[-2]{*}{250}   & \multirow{2}[-2]{*}{6.826} & \multirow{2}[-2]{*}{1.64}  & \multirow{2}[-2]{*}{6.83}  & \multirow{2}[-2]{*}{1.37} \\
          &  \multicolumn{1}{c|}{\multirow{2}[-2]{*}{FP-16}} & \multirow{2}[-2]{*}{5}     & \multirow{2}[-2]{*}{5}     & \multirow{2}[-2]{*}{2}     & \multirow{2}[-2]{*}{13}    & \multirow{2}[-2]{*}{256}   & \multirow{2}[-2]{*}{2.492} & \multirow{2}[-2]{*}{4.48}  & \multirow{2}[-2]{*}{2.49}  & \multirow{2}[-2]{*}{3.92} \\
          &  \multicolumn{1}{c|}{\multirow{2}[-2]{*}{FP-8}} & \multirow{2}[-2]{*}{3}     & \multirow{2}[-2]{*}{3}     & \multirow{2}[-2]{*}{2}     & \multirow{2}[-2]{*}{11}    & \multirow{2}[-2]{*}{300}   & \multirow{2}[-2]{*}{2.108} & \multirow{2}[-2]{*}{5.30}  & \multirow{2}[-2]{*}{2.11}  & \multirow{2}[-2]{*}{5.34} \\
    \bottomrule
    \end{tabular}%
    }
  \label{HDL}%
\end{table}%
\raggedbottom
Table-\ref{compare} compares our LA with other LAs on FPGA. Our HDL designs here are for the highest level of parallelism achieved by the platforms for FP-16. Since the models are not the same, all the performance parameters such as frequency, latency,  throughput, and normalized throughput are measured for a fair comparison. We achieved the lowest latency of 1.42 $\mu$S and the highest GOPS of 7.87 with the HDL design on U55C board running at 250 MHz frequency. Among all of our HDL designs, it exploits full parallelism and gives the highest GOPS/LUT and GOPS/DSP. Work\cite{relatedWork7} achieved latency closest to ours on VC707 at 140 MHz, but its GOPS is $1.73\times$ lower. The LA in \cite{relatedWork10} got $1.5\times$ more GOPS/LUT but $3.5\times$ less GOPS than ours because it consumed fewer LUTs. The HDL design in \cite{relatedWork12} has the highest GOPS/DSP meaning it uses fewer DSPs. While its GOPS is comparable to ours, our slowest HDL design on ZCU104 and our slowest HLS design on VC707 are, respectively, $3.78\times$ and $1.26\times$ faster than this. The HLS design performed better on ZCU104 with the lowest latency and the highest GOPS of 2.92 $\mu$S and 3.83 respectively. Since the design consumes fewer resources, the GOPS/LUT and GOPS/DSP are higher than that of the HDL design. Both our HDL and HLS designs are respectively $280\times$ and $136\times$ faster than the ARM Core CPU running at 1.2GHz frequency.
\begin{table}[htbp]
\setlength{\arrayrulewidth}{0.2mm}
\renewcommand{\arraystretch}{1.5}
  \centering
  \caption{Comparison with Other LSTM Accelerators}
  \resizebox{0.8\columnwidth}{3.5cm}{%
    \begin{tabular}{|c|c|c|c|c|c|c|c|}
    \toprule
    \multirow{2}[1]{*}{Work} & \multirow{2}[1]{*}{Platform} & \multirow{2}[1]{*}{Method} & Fmax & Latency & Throughput  & GOPS/ & GOPS/ \bigstrut[t]\\
      &   &   & (MHz) & ($\mu$S) & (GOPS) & (LUT*1000) & (DSP*1000000) \\
    \midrule
    \multicolumn{1}{|c|}{\cite{relatedWork1}} & \multicolumn{1}{c|}{VC707} & HLS & 150 & 390 & 7.26 & 38.23 & 6.17 \bigstrut[b]\\
    \cite{relatedWork2} & \multicolumn{1}{c|}{VC707} & HLS & 150 & 4.3 & 13.45 & 47 & 7.77 \bigstrut[t]\\
    \cite{relatedWork3} & U250 & HLS & \cellcolor[rgb]{ 1,  .851,  .4}300 & \cellcolor[rgb]{ 1,  .851,  .4}0.867 & \multicolumn{1}{>{\columncolor[rgb]{ 1,  .851,  .4}}c|}{17.2} & -- & 1.9 \bigstrut\\
    \cite{relatedWork8} & Zynq-7020 & HLS & 118 & 18760 & 0.00977 & 1.14 & 0.143 \bigstrut\\
    \cite{relatedWork9} & Artix-7 & HDL & 160 & 800 & 0.631 & -- & -- \bigstrut\\
    \cite{relatedWork11} & Artix-7 & HDL & 53 & 1240 & 0.055 & 56 & 13.75 \bigstrut\\
    \cite{relatedWork10} & XC7Z030 & HDL & 100 & -- & 2.26 & \cellcolor[rgb]{ 1,  .851,  .4}98.1 & -- \bigstrut\\
        \multicolumn{1}{|c|}{\cite{relatedWork7}} & \multicolumn{1}{c|}{VC707} & HDL & 140 & 2.05 & 4.535 & 31.2 & 5.06 \bigstrut\\
    \cite{relatedWork12} & XC7Z020 & HDL & 164 & 9.3 & 7.51 & -- & \cellcolor[rgb]{ 1,  .851,  .4}192 \bigstrut[b]\\
    \multicolumn{1}{|c|}{\cite{relatedWork6}} & \multicolumn{1}{c|}{ZC7020} & -- & 142 & 932 & 1.049 & 16.96 & -- \bigstrut\\
    \midrule
\hline
    
    \multicolumn{1}{|c|}{\multirow{8}[-6]{*}{This}} & \multicolumn{1}{c|}{ARM Cortex} & Embedded & \multirow{3}[-2]{*}{1200} & \multirow{3}[-2]{*}{398} & \multirow{3}[-2]{*}{0.028} & \multirow{3}[-2]{*}{--}  & \multirow{3}[-2]{*}{--}  \\ 
    
    \multicolumn{1}{|c|}{\multirow{8}[-6]{*}{Work}} & \multicolumn{1}{c|}{A53} & C &  &  &  &  &  \bigstrut\\ 
    \hhline{~|*{7}{-}}
     & U55C & \multirow{3}[6]{*}{HDL} & \multicolumn{1}{|c|}{\cellcolor[rgb]{ .973,  .796,  .678}250} & \cellcolor[rgb]{ .973,  .796,  .678}1.42 & \cellcolor[rgb]{ .973,  .796,  .678}7.87 & \cellcolor[rgb]{ .973,  .796,  .678}65.67 & \cellcolor[rgb]{ .973,  .796,  .678}3.9 \bigstrut\\
     & ZCU104 &   & 215 & 2.46 & 4.56 & 67 & 3.99 \bigstrut\\
& VC707 &   & 166 & 2.06 & 5.37 & 45.5 & 2.67 \bigstrut\\ \hhline{~|*{7}{-}}
& U55C & \multirow{3}[6]{*}{HLS} & \cellcolor[rgb]{ .973,  .796,  .678}375 & 4.72 & 2.36 & 93.42 & 10.57 \bigstrut\\ 
& ZCU104 &   & 350 & \cellcolor[rgb]{ .973,  .796,  .678}2.92 & \cellcolor[rgb]{ .973,  .796,  .678}3.83 & \cellcolor[rgb]{ .973,  .796,  .678}105 & \cellcolor[rgb]{ .973,  .796,  .678}17 \bigstrut\\
& VC707 &   & 213 & 7.40 & 1.51 & 49.45 & 6.7 \bigstrut\\
    \bottomrule
    \end{tabular}%
    }
  \label{compare}%
\end{table}%


\section{Conclusion}\label{conclude}
In this research, we demonstrated a custom LSTM accelerator on FPGA created using both high-level synthesis (HLS) and hardware description language (HDL). For a use case involving high-rate time series data that were dynamically generated by simulating a ballistic environment, we created a new three-layer LSTM model. To address the demands for real-time reaction, the hardware accelerator for this model was subsequently constructed on an FPGA. Even with the use of certain directives, the HLS design process produces an unmanageable circuit despite being quick and simple to modify. On the other hand, while the HDL design process is lengthy, it can still result in the intended circuit, and we were able to manage the degree of parallelism. We contrasted the two designs' performance, utility, and adaptability. Then, we analyzed the differences between the performance, utilization, and flexibility of the two design strategies. The ZCU104 platform uses the outermost loop pipelining pragma to provide the lowest latency for HLS design. The outermost loop unrolling pragma can use more resources (DSP) in HLS, but it did not achieve latency that was lower than the outermost loop pipelining pragma. High resource usage may be enabled for the HDL design. As a result, we could set up the U55C so that it can fully parallelize our LSTM model, which has the highest DSP usage. It had the lowest latency at full parallelism as a consequence. Yet, ZCU104 also outperformed U55C in HDL design at the same amount of parallelism. Our HLS and HDL designs are significantly faster than the CPU, according to experimental findings. In terms of latency, throughput, frequency, or normalized throughput, the findings further demonstrate that our approach is better than the majority of current LSTM accelerators on FPGA.

\vspace{12pt}
\color{red}


\begin{thebibliography}{10}
\providecommand{\url}[1]{#1}
\csname url@samestyle\endcsname
\providecommand{\newblock}{\relax}
\providecommand{\bibinfo}[2]{#2}
\providecommand{\BIBentrySTDinterwordspacing}{\spaceskip=0pt\relax}
\providecommand{\BIBentryALTinterwordstretchfactor}{4}
\providecommand{\BIBentryALTinterwordspacing}{\spaceskip=\fontdimen2\font plus
\BIBentryALTinterwordstretchfactor\fontdimen3\font minus
  \fontdimen4\font\relax}
\providecommand{\BIBforeignlanguage}[2]{{%
\expandafter\ifx\csname l@#1\endcsname\relax
\typeout{** WARNING: IEEEtran.bst: No hyphenation pattern has been}%
\typeout{** loaded for the language `#1'. Using the pattern for}%
\typeout{** the default language instead.}%
\else
\language=\csname l@#1\endcsname
\fi
#2}}
\providecommand{\BIBdecl}{\relax}
\BIBdecl

\bibitem{lstm1}
T.~Mikolov, M.~Karafi{\'a}t, L.~Burget, J.~H. {\vC}ernock{\'y}, and
  S.~Khudanpur, ``Recurrent neural network based language model,'' in
  \emph{Interspeech}, 2010.

\bibitem{lstm2}
W.~Byeon, T.~M. Breuel, F.~Raue, and M.~Liwicki, ``Scene labeling with lstm
  recurrent neural networks,'' in \emph{2015 IEEE Conference on Computer Vision
  and Pattern Recognition (CVPR)}, 2015, pp. 3547--3555.

\bibitem{Nelson2022GeneratedDatasetsDynamic}
M.~Nelson, S.~Laflamme, C.~Hu, A.~G. Moura, J.~Hong, A.~Downey, P.~Lander,
  Y.~Wang, E.~Blasch, and J.~Dodson, ``Generated datasets from dynamic
  reproduction of projectiles in ballistic environments for advanced research
  ({DROPBEAR}) testbed,'' \emph{{IOP} {SciNotes}}, vol.~3, no.~4, p. 044401,
  nov 2022.

\bibitem{high-rate4}
\BIBentryALTinterwordspacing
\emph{{Progress Towards Data-Driven High-Rate Structural State Estimation on
  Edge Computing Devices}}, ser. International Design Engineering Technical
  Conferences and Computers and Information in Engineering Conference, vol.
  Volume 10: 34th Conference on Mechanical Vibration and Sound (VIB), 08 2022,
  v010T10A017. [Online]. Available:
  \url{https://doi.org/10.1115/DETC2022-90118}
\BIBentrySTDinterwordspacing

\bibitem{high-rate6}
\BIBentryALTinterwordspacing
A.~Downey, J.~Hong, J.~Dodson, M.~Carroll, and J.~Scheppegrell, ``Millisecond
  model updating for structures experiencing unmodeled high-rate dynamic
  events,'' \emph{Mechanical Systems and Signal Processing}, vol. 138, p.
  106551, 2020. [Online]. Available:
  \url{https://www.sciencedirect.com/science/article/pii/S0888327019307721}
\BIBentrySTDinterwordspacing

\bibitem{Hong2018IntroductionStateEstimation}
J.~Hong, S.~Laflamme, J.~Dodson, and B.~Joyce, ``Introduction to state
  estimation of high-rate system dynamics,'' \emph{Sensors}, vol.~18, no.~2, p.
  217, jan 2018.

\bibitem{high-rate1}
J.~Hong, S.~Laflamme, and J.~Dodson, ``Study of input space for state
  estimation of high-rate dynamics,'' \emph{Structural Control and Health
  Monitoring}, vol.~25, no.~6, p. e2159, 2018.

\bibitem{high-rate2}
\BIBentryALTinterwordspacing
M.~Nelson, V.~Barzegar, S.~Laflamme, C.~Hu, A.~R. Downey, J.~D. Bakos,
  A.~Thelen, and J.~Dodson, ``Multi-step ahead state estimation with hybrid
  algorithm for high-rate dynamic systems,'' \emph{Mechanical Systems and
  Signal Processing}, vol. 182, p. 109536, 2023. [Online]. Available:
  \url{https://www.sciencedirect.com/science/article/pii/S0888327022006379}
\BIBentrySTDinterwordspacing

\bibitem{high-rate3}
\BIBentryALTinterwordspacing
\emph{{Microsecond State Monitoring of Nonlinear Time-Varying Dynamic
  Systems}}, ser. Smart Materials, Adaptive Structures and Intelligent Systems,
  vol. Volume 2: Modeling, Simulation and Control of Adaptive Systems;
  Integrated System Design and Implementation; Structural Health Monitoring, 09
  2017, v002T05A013. [Online]. Available:
  \url{https://doi.org/10.1115/SMASIS2017-3999}
\BIBentrySTDinterwordspacing

\bibitem{high-rate5}
J.~Dodson, A.~Downey, S.~Laflamme, M.~D. Todd, A.~G. Moura, Y.~Wang, Z.~Mao,
  P.~Avitabile, and E.~Blasch, ``High-rate structural health monitoring and
  prognostics: An overview,'' in \emph{Data Science in Engineering, Volume 9},
  R.~Madarshahian and F.~Hemez, Eds.\hskip 1em plus 0.5em minus 0.4em\relax
  Cham: Springer International Publishing, 2022, pp. 213--217.

\bibitem{data}
\BIBentryALTinterwordspacing
High-Rate-SHM-Working-Group, ``Acceleration-vs-roller-displacement dataset for
  dropbear.'' [Online]. Available:
  \url{https://github.com/High-Rate-SHM-Working-Group/Dataset-2-DROPBEAR-Acceleration-vs-Roller-Displacement}
\BIBentrySTDinterwordspacing

\bibitem{testbed}
B.~Joyce, J.~Dodson, S.~Laflamme, and J.~Hong, ``An experimental test bed for
  developing high-rate structural health monitoring methods,'' \emph{Shock and
  Vibration}, vol. 2018, 2018.

\bibitem{atiyeh}
A.~Panahi, E.~Kabir, A.~Downey, D.~Andrews, M.~Huang, and J.~D. Bakos,
  ``High-rate machine learning for forecasting time-series signals,'' in
  \emph{2022 IEEE 30th Annual International Symposium on Field-Programmable
  Custom Computing Machines (FCCM)}, 2022, pp. 1--9.

\bibitem{relatedWork1}
Y.~Guan, Z.~Yuan, G.~Sun, and J.~Cong, ``Fpga-based accelerator for long
  short-term memory recurrent neural networks,'' in \emph{2017 22nd Asia and
  South Pacific Design Automation Conference (ASP-DAC)}.\hskip 1em plus 0.5em
  minus 0.4em\relax IEEE, 2017, pp. 629--634.

\bibitem{relatedWork2}
Z.~Sun, Y.~Zhu, Y.~Zheng, H.~Wu, Z.~Cao, P.~Xiong, J.~Hou, T.~Huang, and
  Z.~Que, ``Fpga acceleration of lstm based on data for test flight,'' in
  \emph{2018 IEEE International Conference on Smart Cloud (SmartCloud)}, 2018,
  pp. 1--6.

\bibitem{relatedWork3}
Z.~Que, E.~Wang, U.~Marikar, E.~Moreno, J.~Ngadiuba, H.~Javed,
  B.~Borzyszkowski, T.~Aarrestad, V.~Loncar, S.~Summers, M.~Pierini, P.~Y.
  Cheung, and W.~Luk, ``Accelerating recurrent neural networks for
  gravitational wave experiments,'' in \emph{2021 IEEE 32nd International
  Conference on Application-specific Systems, Architectures and Processors
  (ASAP)}, 2021, pp. 117--124.

\bibitem{relatedWork8}
\BIBentryALTinterwordspacing
U.~Yoshimura, T.~Inoue, A.~Tsuchiya, and K.~Kishine,
  ``\BIBforeignlanguage{en}{Implementation of {Low}-{Energy} {LSTM} with
  {Parallel} and {Pipelined} {Algorithm} in {Small}-{Scale} {FPGA}},'' in
  \emph{\BIBforeignlanguage{en}{2021 {International} {Conference} on
  {Electronics}, {Information}, and {Communication} ({ICEIC})}}.\hskip 1em plus
  0.5em minus 0.4em\relax Jeju, Korea (South): IEEE, Jan. 2021, pp. 1--4.
  [Online]. Available: \url{https://ieeexplore.ieee.org/document/9369806/}
\BIBentrySTDinterwordspacing

\bibitem{relatedWork4}
Q.~Liu, T.~Liang, Z.~Huang, and V.~Dinavahi, ``Real-time fpga-based hardware
  neural network for fault detection and isolation in more electric aircraft,''
  \emph{IEEE Access}, vol.~7, pp. 159\,831--159\,841, 2019.

\bibitem{relatedWork5}
L.~G. Rocha, M.~Liu, D.~Biswas, B.-E. Verhoef, S.~Bampi, C.~H. Kim,
  C.~Van~Hoof, M.~Konijnenburg, M.~Verhelst, and N.~V. Helleputte, ``Real-time
  hr estimation from wrist ppg using binary lstms,'' in \emph{2019 IEEE
  Biomedical Circuits and Systems Conference (BioCAS)}, 2019, pp. 1--4.

\bibitem{relatedWork9}
\BIBentryALTinterwordspacing
A.~N. Mazumder, H.-A. Rashid, and T.~Mohsenin, ``\BIBforeignlanguage{en}{An
  {Energy}-{Efficient} {Low} {Power} {LSTM} {Processor} for {Human} {Activity}
  {Monitoring}},'' in \emph{\BIBforeignlanguage{en}{2020 {IEEE} 33rd
  {International} {System}-on-{Chip} {Conference} ({SOCC})}}.\hskip 1em plus
  0.5em minus 0.4em\relax Las Vegas, NV, USA: IEEE, Sep. 2020, pp. 54--59.
  [Online]. Available: \url{https://ieeexplore.ieee.org/document/9524796/}
\BIBentrySTDinterwordspacing

\bibitem{relatedWork11}
N.~K. Manjunath, H.~Paneliya, M.~Hosseini, D.~Hairston, and T.~Mohsenin,
  ``\BIBforeignlanguage{en}{A {Low}-{Power} {LSTM} {Processor} for
  {Multi}-{Channel} {Brain} {EEG} {Artifact} {Detection}}.''

\bibitem{relatedWork14}
\BIBentryALTinterwordspacing
J.~Yin, J.~Han, R.~Xie, C.~Wang, X.~Duan, Y.~Rong, X.~Zeng, and J.~Tao,
  ``\BIBforeignlanguage{en}{{MC}-{LSTM}: {Real}-{Time} {3D} {Human} {Action}
  {Detection} {System} for {Intelligent} {Healthcare} {Applications}},''
  \emph{\BIBforeignlanguage{en}{IEEE Transactions on Biomedical Circuits and
  Systems}}, vol.~15, no.~2, pp. 259--269, Apr. 2021. [Online]. Available:
  \url{https://ieeexplore.ieee.org/document/9373938/}
\BIBentrySTDinterwordspacing

\bibitem{comparative1}
R.~Mill{\'o}n, E.~Frati, and E.~Rucci, ``A comparative study between hls and
  hdl on soc for image processing applications,'' \emph{arXiv preprint
  arXiv:2012.08320}, 2020.

\bibitem{comparative2}
H.~S. Lee and J.~W. Jeon, ``Comparison between hls and hdl image processing in
  fpgas,'' in \emph{2020 IEEE International Conference on Consumer Electronics
  - Asia (ICCE-Asia)}, 2020, pp. 1--2.

\bibitem{microblaze}
\BIBentryALTinterwordspacing
``\BIBforeignlanguage{en}{{MicroBlaze} {Processor} {Reference} {Guide}},''
  2022. [Online]. Available:
  \url{https://docs.xilinx.com/v/u/en-US/ug984-vivado-microblaze-ref}
\BIBentrySTDinterwordspacing

\bibitem{DMA}
\BIBentryALTinterwordspacing
``Introduction • {DMA}/{Bridge} {Subsystem} for {PCI} {Express} {Product}
  {Guide} ({PG195}) • {Reader} • {Documentation} {Portal}.'' [Online].
  Available: \url{https://docs.xilinx.com/r/en-US/pg195-pcie-dma}
\BIBentrySTDinterwordspacing

\bibitem{CNN-MLPA}
E.~Kabir, A.~Poudel, Z.~Aklah, M.~Huang, and D.~Andrews, ``A runtime
  programmable accelerator for convolutional and multilayer perceptron neural
  networks on fpga,'' in \emph{Applied Reconfigurable Computing. Architectures,
  Tools, and Applications: 18th International Symposium, ARC 2022, Virtual
  Event, September 19–20, 2022, Proceedings}.\hskip 1em plus 0.5em minus
  0.4em\relax Berlin, Heidelberg: Springer-Verlag, 2022, p. 32–46.

\bibitem{relatedWork7}
\BIBentryALTinterwordspacing
J.~C. Ferreira and J.~Fonseca, ``\BIBforeignlanguage{en}{An {FPGA}
  implementation of a long short-term memory neural network},'' in
  \emph{\BIBforeignlanguage{en}{2016 {International} {Conference} on
  {ReConFigurable} {Computing} and {FPGAs} ({ReConFig})}}.\hskip 1em plus 0.5em
  minus 0.4em\relax Cancun, Mexico: IEEE, Nov. 2016, pp. 1--8. [Online].
  Available: \url{http://ieeexplore.ieee.org/document/7857151/}
\BIBentrySTDinterwordspacing

\bibitem{relatedWork10}
E.~Azari and S.~Vrudhula, ``An {Energy}-{Efficient} {Reconfigurable} {LSTM}
  {Accelerator} for {Natural} {Language} {Processing},'' in \emph{2019 {IEEE}
  {International} {Conference} on {Big} {Data} ({Big} {Data})}, Dec. 2019, pp.
  4450--4459.

\bibitem{relatedWork12}
\BIBentryALTinterwordspacing
E.~Bank-Tavakoli, S.~A. Ghasemzadeh, M.~Kamal, A.~Afzali-Kusha, and M.~Pedram,
  ``\BIBforeignlanguage{en}{{POLAR}: {A} {Pipelined}/{Overlapped}
  {FPGA}-{Based} {LSTM} {Accelerator}},'' \emph{\BIBforeignlanguage{en}{IEEE
  Transactions on Very Large Scale Integration (VLSI) Systems}}, vol.~28,
  no.~3, pp. 838--842, Mar. 2020. [Online]. Available:
  \url{https://ieeexplore.ieee.org/document/8889770/}
\BIBentrySTDinterwordspacing

\bibitem{relatedWork6}
\BIBentryALTinterwordspacing
A.~X.~M. Chang and E.~Culurciello, ``\BIBforeignlanguage{en}{Hardware
  accelerators for recurrent neural networks on {FPGA}},'' in
  \emph{\BIBforeignlanguage{en}{2017 {IEEE} {International} {Symposium} on
  {Circuits} and {Systems} ({ISCAS})}}.\hskip 1em plus 0.5em minus 0.4em\relax
  Baltimore, MD, USA: IEEE, May 2017, pp. 1--4. [Online]. Available:
  \url{http://ieeexplore.ieee.org/document/8050816/}
\BIBentrySTDinterwordspacing

\end{thebibliography}
\end{document}